# Measurement and Statistical Analysis of End User Satisfaction with Mobile Network Coverage in Afghanistan

M. A. Habibi[1], M. Ulman[2], B. Baha[3], M. Stočes[2]

[1] Institute for Wireless Communication and Navigation, Department of Electrical and Computer Engineering, Technical University of Kaiserslautern, Germany

[2] Department of Information Technologies, Faculty of Economics and Management, Czech University of Life Sciences Prague, Czech Republic

[3] School of Computing, Engineering and Mathematics, Faculty of Science and Engineering, University of Brighton, United Kingdom

## Abstract

Network coverage is one of the fundamental requirements of any business of a service provider. Mobile operators are expected to deploy base stations in an effective way in order to cover most of the residential areas of a particular country. Improved network coverage leads to increase total revenue, provide end users with enhanced Quality of Services (QoS) anytime anywhere, and play vital role in the development of telecom sector. In this paper, we measure and statistically analyze network coverage of mobile operators in Afghanistan. Our study is based on primary data collected on random basis from 1,515 mobile phone users of cellular operators. The relationship between "No Network Coverage" in some residential areas and "Satisfaction of Mobile Phone Users" is also investigated. We furthermore propose realistic, feasible and cost-efficient solutions to mobile operators and policy makers in order to expand network coverage to non-covered residential areas as well as enhance the performance of networks in existing covered areas of the country.

## Keywords

Mobile network, network coverage, measurement, statistical analysis, end user, quality of service, end user satisfaction, Afghanistan.

## Introduction

Mobile network coverage is considered one of the fundamental requirements of cellular networks. In order for an operator to provide coverage for a desired area needs to install base stations in appropriate locations and furthermore optimize network from time to time. The more an operator expands coverage area the better it builds a strong brand, which leads to inspire customer loyalty and grows market share. However, it takes most operators six to seven years to build a network with nation-wide coverage when introducing new technologies. For example, a mobile operator in Germany (Telefonica – which uses the O2 brand) began deploying Long Term Evolution (LTE) in 2011, and it took seven years (2017) to cover more than 95 % of the population. In reality, this operator achieved nation-wide Global System for Mobile Communication (GSM) coverage by using 900 MHz several years ago. They could not however deploy LTE on this frequency as the spectrum was occupied by GSM (HUAWEI, 2017).

Providing of network coverage in urban areas is a challenging task due to high-rise buildings, trees, massive number of end users, etc. Dense located buildings in urban areas do not only attenuate the received signal power but also weaken the undesired signal, i.e., the interference. On the other hand, low-density population in rural areas are not cost-efficient for operators to provide services. However, every residential area is required to be covered and people need to be provided with telecom services equally without considering their locations. Therefore, each area needs its own specific scheme to be covered considering demographic data, QoS, Capital Expenditures (CAPEX), and Operational Expenditures (OPEX).

Recent statistics provided by ITU shows that globally cellular networks cover 95 % of residential areas in the world (i.e. around seven billion people) (ITU, 2016). Mobile broadband networks (3G or above) reach 84 % of the global population but only 67 % of the rural population. By the end of 2016, 3.9 billion people – 53 % of the world's population – was not using the Internet, it specifically means that one out of two people (47 %) in the world are using the Internet. Based on these facts, there are still more residential areas (both rural and urban) in the world, which are required to be covered and people need access to enhanced telecom and internet services as a basic human right (ITU, 2016).

On the other hand, according to (Cisco, 2017), there will be 11.5 billion mobile devices by 2019. Mobile data traffic is globally expected to grow up to 24.3 Exabytes (EB) per month by 2019 – nearly a tenfold increase in comparison to 2014. The ever-increasing amount of mobile data traffic moves technical experts forward to seek various techniques and schemes in order to handle challenges in different parts of mobile networks.

The Afghanistan Telecom Regulatory Authority (ATRA) and Ministry of Communication and Information Technology (MCIT) claim that 89 % of residential areas in the country is covered by network coverage (ATRA, 2016). However, no detailed report has been published to prove and verify the accuracy of this claim. Subsequently, no study has been conducted so far to measure the quality and satisfaction of end user with network coverage. Therefore, in this study we are going to measure and furthermore statistically analyze end user satisfaction with mobile network coverage in Afghanistan. We will also focus on how to improve existing network coverage and furthermore expand it to non-covered residential areas. It is worth noting that, we are only dealing with expansion of network coverage to non-covered areas and improving of outdoor coverage, thus, the optimization of in-door coverage is not in the scope of our study.

There are couple of studies, which empirically analyze end user satisfaction of mobile operators with employing of different methodologies and various variables. A study has found in Turkey that satisfaction of mobile phone user is associated with wide and improved network coverage, efficient customer service, enhanced QoS, and fulfilling the expectations of end users (Aydin and Ozer, 2005).

Three more empirical research studies in Hong Kong (Woo and Fock, 1999), China (Wang and Lo, 2002) and South Korea (Kim and Jeong, 2004) have found many factors i.e. wide and improved network coverage, reasonable pricing policies, enhanced QoS (both voice and data), value added services and customer support which influence end user satisfaction.

A study has recently been conducted in Pakistan discovered that service quality, price rate, brand image, sale promotion and improved network coverage have significant impact on end user satisfaction (Iqbal, 2016). While, (Khan, 2010) has found that tangibles, reliability, responsiveness, convenience, assurance, empathy, and network quality have statistical significance relationship with end users satisfaction.

Moreover, a comprehensive survey based research has recently been conducted to measure and analyze satisfaction of end users with QoS of mobile operators in Afghanistan (Habibi et al., 2016). This study analyzes the relationship between various variables and end user satisfaction through hypotheses testing and furthermore proposes adequate technical solutions for operators in order to overcome existing challenges in the area of QoS. Most of the above studies and in particular (Habibi et al., 2016) measure and furthermore statistically analyze satisfaction of end users with QoS of mobile networks. But a detailed study in order to measure and analyze end user satisfaction toward network coverage, and the most unwanted situations which mobile phone users experience are missing in the literature. In order to fulfil this gap, we have taken this initiative by conducting a survey-based research in Afghanistan. We have thoroughly analyzed and furthermore tested the relationship between "End User Satisfaction" and "Network Coverage". Despite that, we have proposed adequate and realistic recommendations in order to address existing challenges in the area of network coverage in the country.

## Mobile networks in Afghanistan

A modern, secure, effective and nationwide telecom infrastructure in the country helps to stimulate economic growth, raise living standards of ordinary Afghans and restore the traditional sense of community and common purpose that unites the Afghan people. It furthermore enhances the effectiveness, efficiency and transparency of public sector, improves delivery of social services, and builds a peaceful and unified society. Considering geographic feature of the country, mobile networks play vital role in narrowing of the physical distances that separate Afghan villages and towns and furthermore dramatically improve access to educational opportunities, humanitarian relief efforts, and even e-health and e-business services.

Considering geographical feature of Afghanistan, a telecom sector with above characteristics does not only play vital role in the transformation of country but can also bridge South-Asia with Central-Asia and China with Middle East. The country has the opportunity to take advantage of its unique geographical positon, and to become a hub for regional connectivity and economic exchange. One of the most important drivers for regional connectivity and cooperation in telecom sector is the political commitment of the region. To achieve this, the region requires robust institutional frameworks in order to plan and furthermore implement regional connectivity agenda.

The Afghan government has been putting many efforts in the telecom sector since 2002. For the time being, there are in total five mobile (four private and one stated-owned) and one landline state-owned operators in the country. The Afghan Wireless Communication Company (AWCC), Etisalat – Afghanistan (ETA), Mobile Telecommunication Network – Afghanistan (MTNA), Telecommunication Development Company Afghanistan (TDCA or Roshan) are the private and Salam is the state-owned mobile operators. The state has its own public landline operator namely the Afghan Telecom (AFTEL). A detailed history of telecom sector and deep insight to each of the operators can be found in (Hamdard, 2012), (UN-ESCAP, 2015), (Baharustani, 2013) and (infoDev, 2013). All operators have built their own broadband microwave backbones. A nationwide 3,100 kilometers of optical fiber network has also been established by AFTEL throughout the country (Habibi et al., 2016). So far 7,155 base stations are installed which cover 89 % of the residential areas (ATRA, 2016). The MCIT has purchased the first ever satellite in the history of the country (AfghanSat One) in 2014. China and Afghanistan have signed an agreement on April 20, 2017 to launch the second satellite (AfghanSat Two) to the space and furthermore lay direct 480 kilometers of optical fiber to connect both neighboring countries (3GCA, 2017).

Alongside all these positive changes and development, there are still some challenges within the telecom sector requiring more effort to be overcome. It can be derived from the previous paragraph that mobile operators do not cover 11 % of the residential areas and Afghan citizens living in those areas still do not have access to telephony and internet services. According to (Habibi et al., 2016), 16 % of end users are unsatisfied, 28 % are neutral, and 3 % are very unsatisfied with the QoS of mobile telephony service. The study has found that 32 % of end users were complaining from low signal intensity, 18 % from blocked calls, 17 % from dropped calls, 11 % from echo, and 14 % from noise during usage of mobile telephony (Habibi et al., 2016). The authors in (Habibi et al., 2016) have further discovered that 32 % of end users are unsatisfied and 12 % are very unsatisfied with mobile internet service.

There are still more efforts needed to be put by policy makers and public and private sector in order to overcome the abovementioned challenges, provide fully nationwide telecom services, enhance the QoS and increase end users' satisfaction, and build a modern and secure telecom infrastructure. The ATRA/MCIT should first provide solutions to existing challenges in the sector, and then take a step forward to work on regional connectivity in order for the country to act as a digital bridge within the region.

## Network coverage footprint

A cellular network is distributed over small areas called cells; each cell is served by at least one fixed-location transceiver known as a cell site or base station. The cell site can further be divided into sectors. The combination of cells is called a cluster, which is served by a Base Station Controller (BSC) or Radio Network Controller (RNC). A cellular network can be made of thousands of cells, hundreds of clusters and tens of RNCs.

These sectors, cells, and clusters are joined together in order to provide network coverage over a large geographic area. When a cellular network provides coverage, it enables a large number of UEs to communicate with each other and with fixed landline telephones anywhere and anytime. The base stations provide mobility facility to both low speed and high speed UEs during telephony and data services. The size, shape, and capacity of a cell and network coverage depend on demographic data and natural factors such as mountains, desserts, highway etc. The quality of coverage is measured in terms of location probability. Therefore, the radio propagation conditions have to be predicted

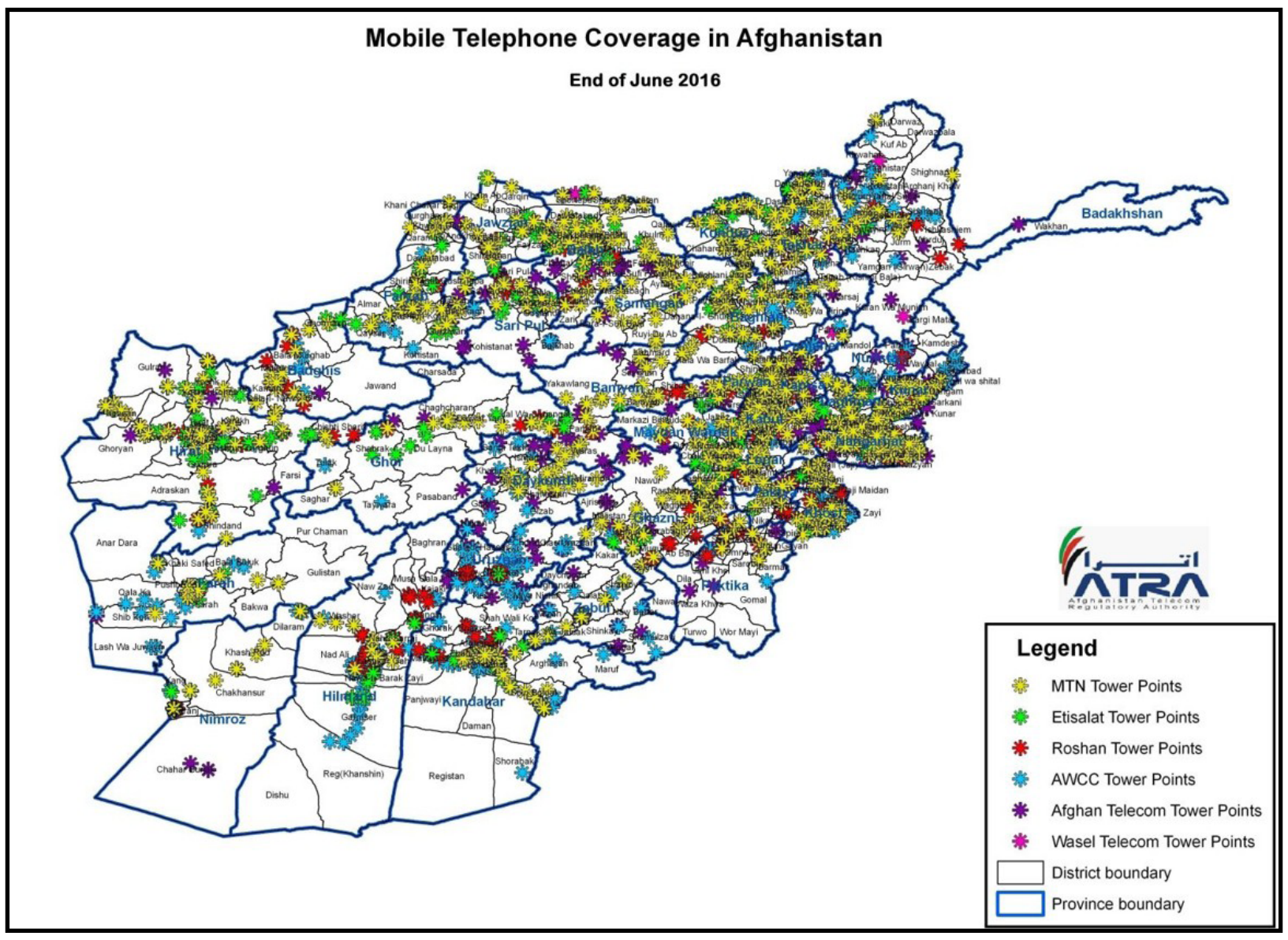

Source: (ATRA, 2016)

Figure 1: Network coverage footprint in Afghanistan.

as accurately as possible for that specific region.

A detailed view of network coverage in Afghanistan provided by 7,155 base stations is shown in Figure 1. The ATRA/MCIT do not publish number of passive and active base stations in the country, so in order for us to predict the accuracy of the network coverage penetration. According to the information we have obtained from some of the engineers workingwith different operators, there are many base stations in different provinces, which are currently switched off by operators due to various reasons. According to their decade of experience, they doubted that mobile networks would cover 89 % of the country. ATRA/MCIT should provide a detailed report of the entire telecom sector every quarter of the year to move operators forward in order to improve network coverage and QoS.

### Existing challenges

Despite all positive developments mentioned in previous section, there are still some challenges on the road to future growth of telecom sector in Afghanistan. Some of these challenges have complex relationship with other sectors i.e. education, security, reconstruction, etc. In this subsection, we will only focus on the existing challenges related to network coverage of mobile operators.

- According to ATRA, cellular networks do not cover 11 % of the residential areas in the country. It means that massive number of Afghans still do not have access to mobile phone and internet services (ATRA, 2016).
- Referring to (Habibi et al, 2016) and the result of our research in this paper, majority of Afghans are not satisfied with the quality of existing 89 % of network coverage. It specifically means that satisfaction of end users has been rarely considered by operators. The quality of existing service and the performance of networks should be improved.
- The continued security threats in some parts of the country present a high level of risk and uncertainty for operators to expand network coverage. The telecom sector itself has become a target from non-state elements. Since February 2008, mobile networks have regularly shut down their base stations at night to avoid attacks from criminal elements, six towers were attacked and five workers have been killed (infoDev, 2013).
- The telecom infrastructure cannot currently support high capacity networks and advanced technology approaches. The optical fiber ring and microwave backbones are not installed according to international norms and standards in some places, the sector is still facing lack of skilled labors, and of course, corruption which is impacting it at certain levels. All of these challenges have direct influence on performance and expansion of network coverage.

In order to overcome abovementioned challenges, the MCIT/ATRA have to come first with modern and strategic solutions to solve these fundamental and infrastructural problems and then move forward to the regional connectivity and implementation of next generation technologies.

## Materials and methods

It is vital to choose an appropriate strategy for measurement of network coverage in order for the operators to have a detailed view of their services and networks across a specific geographic area. Vendors, operators, regulatory bodies, and researchers around the world have been trying to propose different strategies in order to measure network coverage from various perspectives and at different stages of network deployment. Measurements should be chosen corresponding to the needs in the stages of planning, development, installation, and maintenance of network coverage.

There are different ways to measure network coverage i.e. operators measure a specific geographic area using Radio Frequency (RF) test such as drive tests (Aydin and Ozer, 2005) in order to figure out signal strength of propagated RF by their base stations' antennas, regulatory bodies conduct survey for public to find out end users' satisfaction degree from operators' services, and so on.

In this paper, we analyze mobile network coverage from the end user perspective in Afghanistan. We conducted a survey in order to figure out end users' satisfaction with network coverage. The survey of this research was originally conducted to study both network coverage and QoS of mobile networks in Afghanistan, but primary data only related to network coverage is measured and analyzed in this paper. We prepared questionnaire (containing 15 questions) in English, Pashto and Dari languages. All technical terms in the questionnaire were explained in such a way, which were easily understandable for ordinary mobile phone users. The survey only covered end user practices, therefore, we use an effective evaluation method of multiple-choice questionnaire.

## Results and discussion

In the beginning, we conducted a pilot survey on a small group of mobile phone users through in person interviews in Kabul in order to test questionnaire. Based on feedbacks from the target group of respondents, necessary changes have been brought in the strategy as well as confirmed in the final draft of the survey. We have subsequently employed the mix-mode technique of data collection from 1,515 mobile phone users during (August – December) 2015. In practice, we collected 812 respondents over the internet using Google Docs from 30 provinces and volunteer surveyors interviewed 703 respondents in person whithin 14 specific provinces. The total number of respondents who attended the survey from all 30 provinces is shown in Figure 2. The average number of respondents per province (mean) is 50.5.

All respondents have answered about their age, gender, level of education, favorite mobile operator, the time each of them has spent with their favorite mobile operator, and the purpose of mobile phone usage. The result of all these variables are shown in Table 1.

The measured data related to the ages of mobile phone users is shown in Table 1. The result illustrates that more than half of the mobile users are between 18 – 30 years old, which clearly declares generational divide in access and usage of mobile phone service in Afghanistan.

The result of data furthermore finds gender inequality in access to mobile phone service

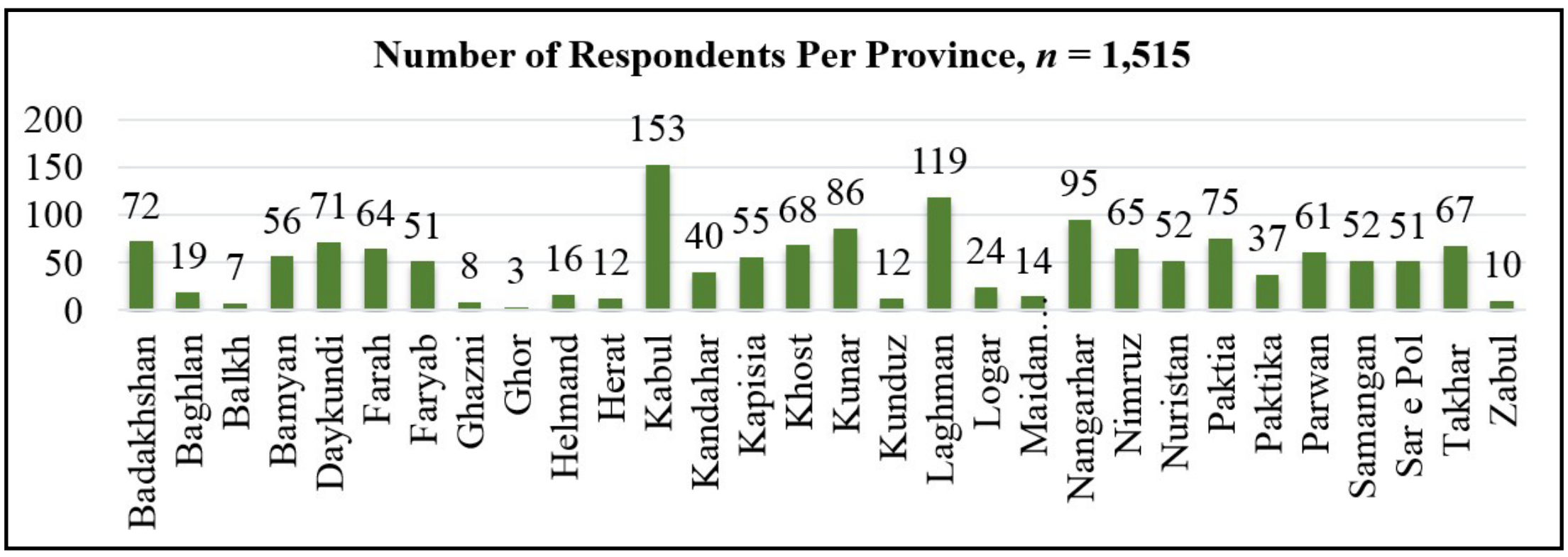

Source: Own processing

Figure 2: Number of respondents per province.

| No. | Variable | Attribute | Outcome (%) | No. | Variable | Attribute | Outcome (%) |
|---|---|---|---|---|---|---|---|
| 1 | Age (Year) | < 18 | 11 | 4 | Operator | AWCC | 14 |
| | | 18 – 30 | 59 | | | ETA | 30 |
| | | 31 – 45 | 24 | | | MTN | 23 |
| | | > 45 | 6 | | | Roshan | 19 |
| 2 | Gender | Male | 75 | | | Salam | 14 |
| | | Female | 25 | 5 | Time with Operator (Years) | < 2 | 28 |
| 3 | Education | Illiterate | 8 | | | 2 – 5 | 52 |
| | | Primary | 13 | | | 6 – 10 | 17 |
| | | Intermediate | 10 | | | > 10 | 3 |
| | | High School | 28 | 6 | Purpose of using of Mobile Phone | Telephony | 48 |
| | | Bachelor | 33 | | | Internet | 4 |
| | | Masters | 8 | | | Both | 48 |
| | | | | | | | |

Source: Own processing

Table 1: General variables of the survey.

in Afghanistan. As shown, three out of four of mobile phone users are males while only 25 % are females. There are couple of reasons of gender inequality in access and usage of mobile phones in a developing country such as Afghanistan. Many research scholars have pointed out that high illiteracy rate of females, patriarchal societies, limited access to telecom services, structural and cultural barriers within the society are main reasons, which led to gender inequality in developing world (Primo, 2003).

Education has significant role in removing barriers on the way to use mobile phone as well as its service (Primo, 2003). Therefore, level of education of end users has been asked in the survey in order to specifically find out that how does significant is it in access and usage of mobile phone service. Based on the results shown in Table 1, only 8 % of end users who use mobile phones are illiterate while the rest of the end users are having various level of education. It is concluded that massive number of mobile phone users in the country are literate while a small amount is illiterate.

The end users were asked about their favorite operator in order to find which one of the operators is leading the market. It can be concluded from the result of the survey shown in Table 1, that ETA is the leading telecom operator in Afghanistan from end user perspective while the state-owned GSM operator Salam is at the end of the list (see the Table 1).

The respondents were further asked about the time they have been using their current SIM card in order to measure end user loyalty. The results in Table 1 shows that roughly half of the end users have been using their favorite network operator from two to five years, while only 3 % stay with their network operator more than ten years. The loyalty of end user depends on various factors, but most important are wider area of network coverage, enhanced QoS, reasonable price for the service, and various types of services offered by the operator.

Mobile phone users use a cellular phone in order to send/receive message, access to internet, send/ receive email, download apps, get directions, recommendations or other location – based information, participate in a video call or video chat, "Check in" or share location, and so on. In the questionnaire, all of the mentioned applications of mobile are divided into 3 categories, mobile for internet purposes, mobile for telephony purposes and mobile for both internet and telephony purposes. The Table 1 furthermore shows, that usage of mobile phone for telephony and internet purposes are roughly same, while only 4 % of mobile phone users in Afghanistan use mobile phone only for internet purposes.

For the time being, the most advanced mobile technology, which is provided by all operators in the country, is UMTS. However, the MCIT has recently announced to launch Fourth Generation (4G) services across the country (Kabultribune, 2017). There are still some areas which are covered by EDGE, GPRS, and in some cases, the end users use Wi-Fi in their homes or small offices. The result of the survey in Figure 3 shows that roughly three out of four of end users use 3G (UMTS) service, 14 % EDGE, 7 % GPRS, and 6 % use mobile internet technologies for Wi-Fi purposes.

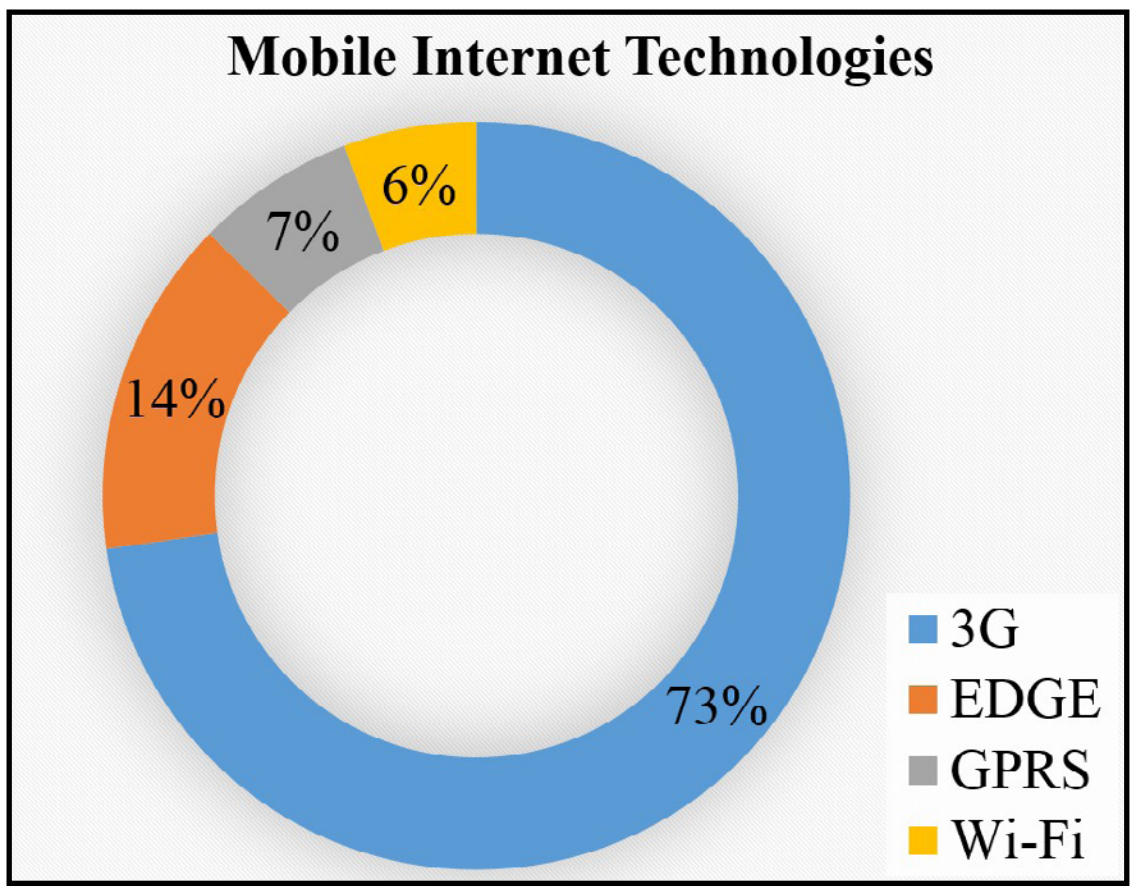

Source: Own processing

Figure 3: Mobile internet technologies.

The survey explores satisfaction degree of end users with mobile network coverage in the country. According to Figure 4, 42.44 % of end users are satisfied with the network coverage of mobile operators in the country, 19.47 % are unsatisfied, 7.32 % are very satisfied, 5.08 % very unsatisfied, and 25.67 % neutral.

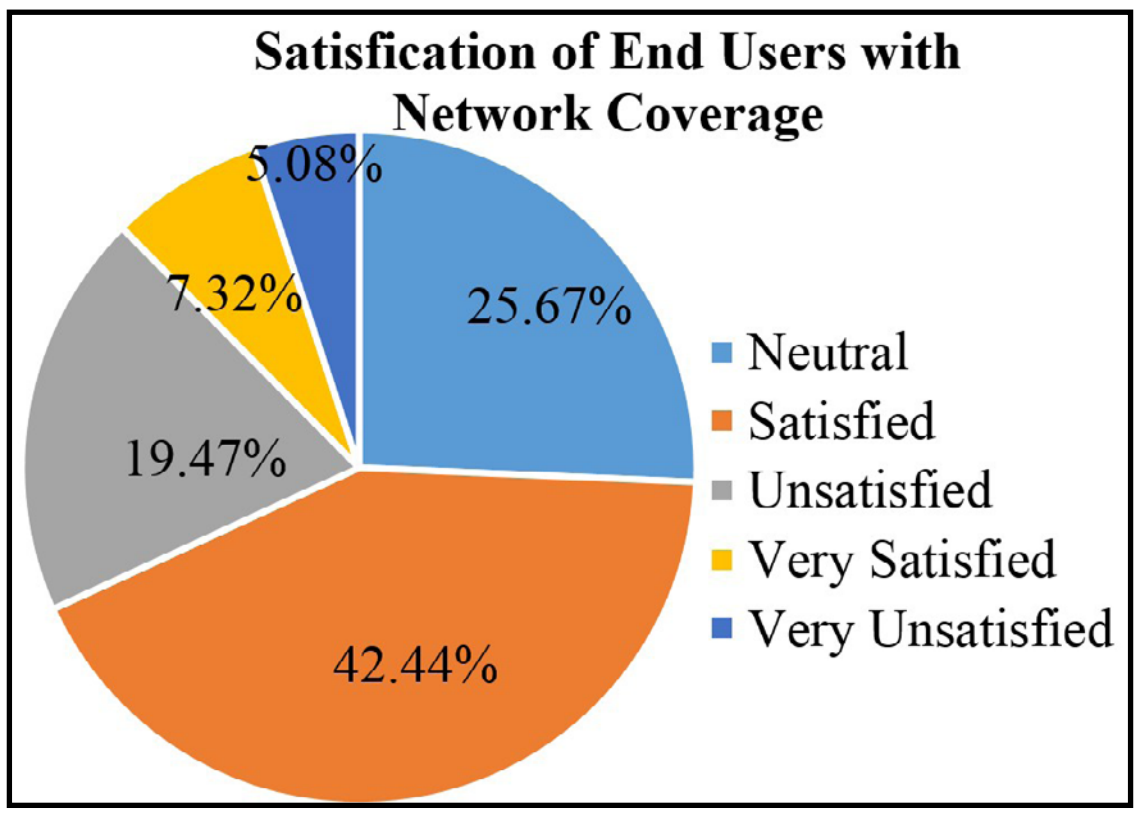

Source: Own processing

Figure 4: Satisfaction of end users with Network Coverage.

The survey furthermore indicates the areas where end users mostly experience "No Network Coverage". The result in Figure 5 shows that most of the end users are experiencing "No Network Coverage" in villages (32.8 %), 8.38 % in the areas, which are closed to country's borders, 13 % in the cities, 25.6 % in the countryside, 15.4 % on highways, the rest 4.6 % of end users are satisfied with network coverage and do not experience "*No Network Coverage*" at any of the above places.

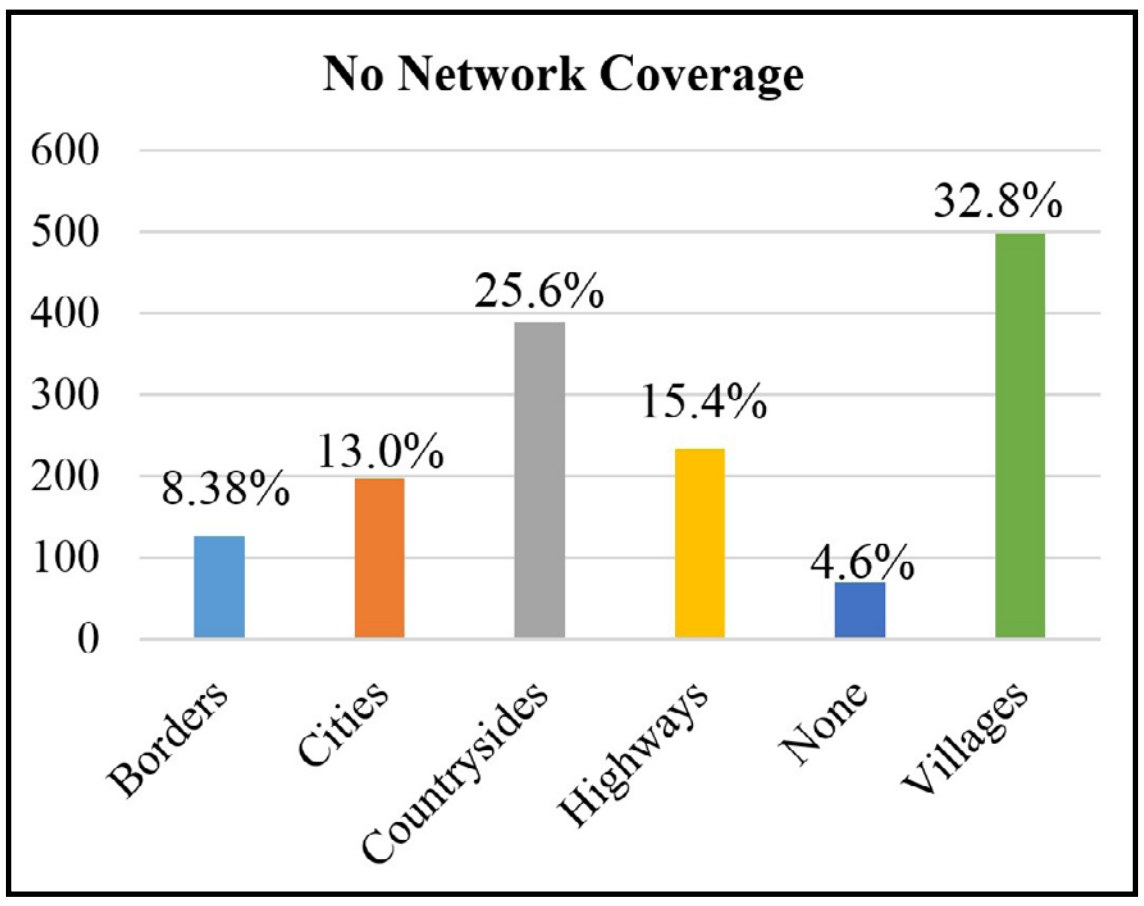

Source: Own processing

Figure 5: Areas where end user experience No Network Coverage.

## Statistical analysis and hypothesis testing

There are in total two categorical variables which create one hypothesis based on the data related to the network coverage of mobile networks in Afghanistan. Associations between these two variables should be tested. Hence, *Goodness-of- Fit (Chi-Square)* test has been chosen to deploy on the hypothesis in order to find dependency.

Before conducting the chi-square test, it is necessary to set up significance level. We have considered 95 % significance level ($\alpha$ = 0.05). As shown in Equation 1, it is determined by multiplying of "number of rows minus one" by "number of columns minus one".

$$DF = (r - 1) * (c - 1) \quad (1)$$

$r$ = No. of rows, $c$ = No. of columns

In next step, the below given formula (Equation 2) is used to perform chi – square test.

$$\chi^2 = \sum_{i=1}^{k} \frac{(O_i - E_i)^2}{E_i} \quad (2)$$

$k$ = No. of categories, $\chi^2$ = Chi-square

$i$ = No. of parameters being estimated

$O_i$ = Observed frequency, $E_i$ = Expected frequency

As mentioned earlier, after obtaining the $\chi^2$ value, it should be compared with critical value from the distribution table considering DF. The final decision is made based on this comparison. If the "chi-square value > table value", the hypothesis is rejected, otherwise, it is impossible to reject.

Out tested hypothesis is stated as follows:

Is there any dependency between categorical variables of 'No Network Coverage in a Geographic Area' and 'Satisfaction of End Users with Network Coverage' of mobile operator in Afghanistan?

The first step is to state the null hypothesis ($H_0$) and alternative hypothesis ($H_1$).

- $H_0$ = There is no association between *'No Network Coverage'* and '*Satisfaction of End Users'*.
- $H_1$ = There is an association between '*Not Network Coverage'* and *'Satisfaction of End Users'*.

The Statistical Application System (SAS) software is used in order to conduct chi-square test, create contingency table and calculate *DF*. Based on the results obtained from SAS, the *DF* and $\chi^2$ values are given below:

$\chi^2 = 307.4957$, $DF = 20$

The contribution table value considering 20 *DF* and $\alpha = 0.05$ is 31.410. As calculated, contribution table value is less than comparing to chi-square test value (307.4957 > 31.410), therefore, the null hypothesis is rejected.

307.4957 > 31.410 > Reject $H_0$.

To conclude, there is statistically significant evidence at $\alpha = 0.05$ that $H_0$ is false. Thus, it can be claimed that, there is dependency between categorical variables of '*No Network Coverage*' and '*Satisfaction of End Users*' of mobile networks in Afghanistan. It specifically means that, *No Network Coverage* in a geographic area has significant impact on the *Satisfaction of End Users* of mobile operators in Afghanistan.

The power of dependency calculated by SAS and shown by *Phi Coefficient (φ)* value is 0.4505, therefore, it can be claimed that there is a high dependency between two categorical variables.

## Proposed solutions and recommendations

There has been considerable research on providing, expanding and optimization of network coverage in all types of areas. However, in real world, the deployment areas of wireless networks are always geographically bounded. It is a much more challenging and significant task to find optimal solutions to cover a specific geographic area. In order to expand network coverage to non-covered residential areas and to improve the quality of existing network coverage in Afghanistan, we propose the following recommendations.

### Inter-operator infrastructure sharing

Afghanistan is made up of rugged mountain terrains, deep valleys and large pieces of deserts. It is very uneconomical and inefficient for a single operator to expand and provide its services for the people who are living in the deserts of Helmand and Kandahar as well as in the mountains of Badakhshan and Kunar. The traditional model of single ownership of all the physical network elements and network layers by mobile network operators has mostly changed in developed countries. Most operators around the world share their infrastructure with others in order to decrease both CAPEX and OPEX. It further results in the expansion of network coverage and improving of QoS, whilst having a very positive environmental impact and optimizing national scarce resources.

In mobile networks, sharing can occur at various levels and different stages. There could be various combinations including the base station site, radio equipment, RAN, Core Network (CN), roaming, power, wireless/wireline backhauls, cooling equipment, antenna cables, antennas, tower masts, and so on. However, all these various kinds of sharing are categorized into three types, passive sharing, active sharing, and roaming-based sharing (Meddour et al., 2011).

The sharing of space or physical supporting infrastructure, which does not require active operational co-ordination between mobile operators' interconnection, is called passive sharing i.e. space or tower sharing. The sharing, which requires operators to share the nodes of active network layers and coordinate with each other during operation, is called active sharing i.e. radio equipment, wireless backhauls and so on. Roaming-based sharing refers to the context of network sharing where an operator relies on the network coverage of another operator and define footprint on either permanent or temporarily basis.

Worldwide, there are many experiences of network infrastructure sharing. At the beginning of 2013, four largest European mobile operators (Deutsche Telecom, France Telecom, Telecom Italia and Telefonica) discussed the possibilities of sharing their resources. The European Commission actively forwarded this cooperation. The occurrence of the new Pan-European operator acting upon network sharing model is possible in the very nearest future due to the efforts of the executive bodies of the European Union (EU) (Thomas and Barker, 2013). Studies have found 90 % cost efficiency between shared and non-shared networks for real dense deployments with both homogenous and heterogeneous infrastructure sharing in Poland (Kibiłda and DaSilva, 2013). Middle Eastern and North African countries i.e. Jordan, Morocco, Oman, Saudi Arabia and the United Arab Emirates mostly have roaming-based sharing infrastructures (Meddour and et al., 2011).

In the light of global experience and considering both cost efficiency and coverage optimization, we strongly recommend to both private and public sectors as well as policy makers in Afghanistan to deploy inter-operator infrastructure sharing in order to expand network coverage for the rest of 11 % of uncovered residential areas, which is going to further result in enhancing of QoS. We encourage the ATRA to take this initiative by extensively studying the experience of some of abovementioned countries/operators and form regulations, which include legal, technical and commercial aspects of inter-operator infrastructure sharing scheme.

### Regular network optimization

Network optimization can help operators to assess and audit the performance of their networks using the full range of available data, and identify various aspects of the design and operation that can be improved. There is certain number of tests and tools which are used to perform network optimization within a specific period of time. The audit will typically result in a series of recommendations and an action plan for network design and performance improvement, along with a process for ongoing performance review and analysis, which is going to reduce outages, improve the customer experience, and simplify network control and operations.

We recommend telecom operators to be strongly committed to the regularity of network optimizations in order to improve network coverage and enhance QoS. We furthermore propose to the ATRA to monitor the regular optimization process of telecom operators and to share the performance report of the entire telecom sector quarterly or at least once a year. These reports are going to move operators forward to enhance the QoS and improve network coverage. Meanwhile, the civil society and mobile phone users will have access to update information and current status of this sector.

### Telecom Development Fund

The MCIT/ATRA have a certain amount of fund namely Telecommunication Development Fund (TDF) under the Universal Access Program (UAP). The main purpose of this fund is to promote rural access to telecom services. In this program, a specific residential area is targeted to be covered by mobile networks, the project is announced and the construction of base stations are awarded to mobile operators on a competitive basis and will be operated and maintained by these operators upon completion. Despite having this fund, the MCIT/ATRA still is not able to expand the coverage to non-covered residential areas or improve the quality of existing network coverage.

We strongly recommend to the policy makers, the MCIT and the ATRA to use this fund efficiently, effectively, and furthermore encourage mobile operators to provide telecom services to non-covered areas.

### Low orbit satellite

Due to the aim to offer mobile and data services in rural residential areas with very low density of end users, some operators developed a solution consisting to use low orbital satellites. However, this approach is not easy to be operated. The main drawback of this solution is the high level of sophistication needed to build, launch and operate a satellite as well as high CAPEX.

In the light of low orbit satellite experience, we also recommend to operators particularly to the state-owned ones to use low orbit satellite in order to provide services for rural areas in some districts such as Bamyan, Daykundi, Badakhsham, Noristan, and other provinces. However, the AFTEL has established a satellite-based network in 800 villages to provide access to the telephony and internet services in rural areas and countryside. This program is called Village Communications Network (VCN). A terminal covers each village in a specific area and the terminal is subsequently connected to the satellite. The MCIT claims that over a million of Afghans have benefited from VCN project (infoDev, 2013). The numbers in Figure 6 represents the locations and the density of terminals in the corresponding area. However, there are still more villages and districts that are not covered and the MCIT/ATRA are expected to do more using this scheme.

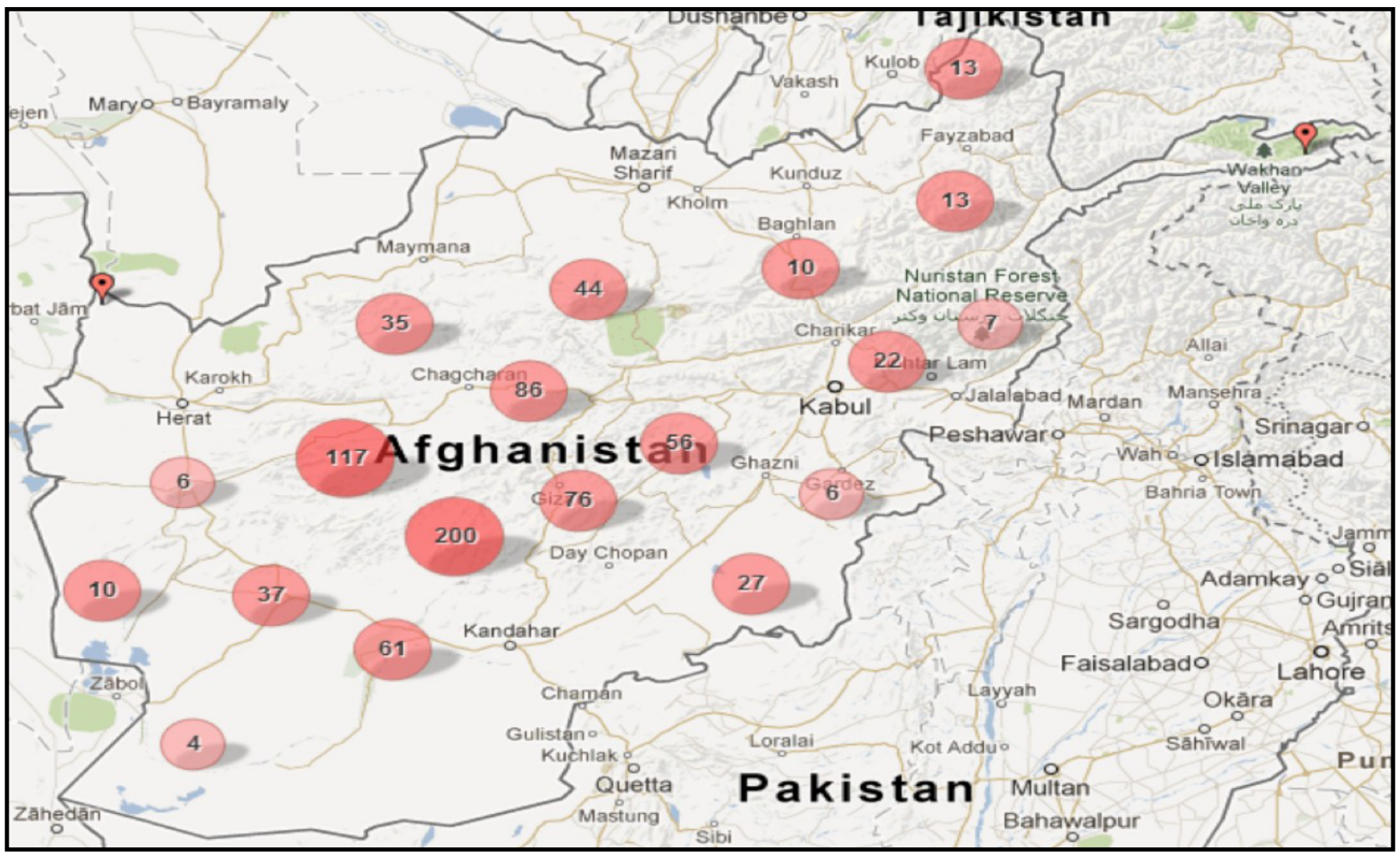

Source: (infoDev, 2013)

Figure 6: Village communication network in Afghanistan.

### Implementation of telecom standardization

Despite above recommendations, one of the most vital factors which increasingly enhance quality of network coverage is the implementation of technical schemes such as deployment of small cells, high order sectorization, infrastructure sharing, beamforming, etc. according to ITU standardization. There is no study to prove that mobile operators and the ATRA mostly or rarely do not follow those unified global standards, but based on field observations of authors from the field, it can be claimed that standardizations are not followed from time to time in the telecom sector of Afghanistan. For example, while deployment of national optical fiber ring, the contractors used to put the cable in depth of few centimeters, which are not according to international standardization. Therefore, we strongly recommend to the MCIT/ATRA to monitor the deployment of global unified standardization in telecom sector in order to enhance the performance of mobile networks, which furthermore result to improve end users' satisfaction.

## Conclusion

In this paper, we measured and statistically analyzed end user satisfaction with network coverage of mobile operators in Afghanistan. We thoroughly studied the current status of network coverage along with existing challenges. We have also tested the relationship between *Satisfaction of Mobile Phone Users* and *No Network Coverage* in some residential areas. In order to expand mobile services to non-covered residential areas and to enhance the quality of network coverage, we have recommended infrastructure sharing, regular network optimization, efficient and effective use of TDF, low orbit satellite and consideration of unified global standards during network deployment as adequate solutions for telecom sector of Afghanistan. In future, we are interested to work on each of the proposed solutions and recommendations of this paper individually.

## Acknowledgements

This work has been performed in the framework of the H2020-ETN-5GAuRA project under grant no. 675806 of the European Commission. The authors would like to acknowledge the contributions of their colleagues. This information reflects the consortium's view, but the consortium is not liable for any use that may be made of any of the information contained therein.

The results and knowledge included herein have been obtained owing to support from the Internal grant agency of the Faculty of Economics and Management, Czech University of Life Sciences in Prague, grant no. 20171005, Animal Behavior Monitoring in Building units".

*Corresponding author:*
*Mohammad Asif Habibi*
*Institute for Wireless Communication and Navigation, Department of Electrical and Computer Engineering, Technical University of Kaiserslautern, 67663 Kaiserslautern, Germany.*
*Phone: +49 152 279 212 06, E-mail: asif@eit.uni-kl.de*